\documentclass[12pt]{article}
\usepackage{graphicx}
\usepackage{amsfonts}
\usepackage{amssymb,amsmath}
\usepackage{latexsym}
\usepackage{color}
\usepackage{cite}
\input{colordvi.tex}

\newcommand{\remb}[1]{{\bf [#1]}}

\begin{document}

\begin{titlepage}

\begin{flushright}
ICRR-Report-579-2010-12
IPMU10-0229
RESCEU-29/10
YITP-10-111 \\
\end{flushright}

\begin{center}

{\Large \bf 
Improved estimation of radiated axions from cosmological axionic strings
}

\vskip .45in

{\large
Takashi Hiramatsu$^1$, Masahiro Kawasaki$^{2,3}$, 
Toyokazu Sekiguchi$^2$, Masahide Yamaguchi$^4$, 
Jun'ichi Yokoyama$^{5,3}$
}

\vskip .45in

{\em
$^1$Yukawa Institute for Theoretical Physics, Kyoto University,
Kyoto 606-8502, Japan\\
$^2$Institute for Cosmic Ray Research, The University of Tokyo,
Kashiwa 277-8582, Japan  \vspace{0.2cm} \\
$^3$Institute for the Physics and Mathematics of the Universe,
The University of Tokyo, Kashiwa 277-8568, Japan \vspace{0.2cm}\\
$^4$Department of Physics, Tokyo Institute of Technology, 
Tokyo 152-8551, Japan \\
$^5$Research Center for the Early Universe,
Graduate School of Science, \\ The University of Tokyo, Tokyo 113-0033, Japan
}

\end{center}

\vskip .4in

\begin{abstract}
Cosmological evolution of axionic string network is analyzed in terms
of field-theoretic simulations in a box of $512^3$ grids, which are
the largest ever, using a new and more efficient identification
scheme of global strings.  The scaling parameter is found to be
$\xi=0.87\pm 0.14$ in agreement with previous results.  The energy
 spectrum is calculated precisely using a pseudo power spectrum 
 estimator which significantly reduces the error in the mean reciprocal
comoving momentum.  The resultant constraint on the axion decay
constant leads to $f_a\le 3\times10^{11}$GeV. 
We also discuss implications for the early Universe.
\if
We estimate the energy spectrum of axions radiated from cosmological
axionic strings. We improve the previous analysis done by Yamaguchi,
Kawasaki and Yokoyama in 1999 (YKY99) based on a field theoretic
simulation.  We develop a new method for identification of global
strings from the field data at discrete grid points.  We also introduce
a pseudo power spectrum estimator to remove the contamination of string
cores in estimation of the energy spectrum.  Applying these new
techniques to our field theoretic simulation with the larger number of
grids, we obtain a precise energy spectrum of radiated
axions. The energy spectrum shows a sharp peak at momentum around
the inverse of horizon scale and exponential suppression toward higher
momenta. Our result shows a good agreement with YKY99 and has smaller 
uncertainties. 
Using our result of numerical
analysis, we obtain a constraint on the decay constant of axion 
$f_a\le 3\times10^{11}$GeV. We also discuss implications for the early Universe.
\fi
\end{abstract}

\end{titlepage}

\setcounter{page}{1}

\section{Introduction} \label{sec:introduction}

The axion has rich implications for astrophysics and cosmology.  
In origin, the
axion is a pseudo Nambu-Goldstone (NG) boson of the Peccei-Quinn (PQ)
symmetry, which is introduced to the Standard Model as a solution to
the Strong CP problem in quantum chromodynamics (QCD)
\cite{Peccei:1977hh}.  Below the QCD scale $\Lambda_\mathrm{QCD}\simeq
200$MeV, the axion acquires a mass from the instanton effect
\cite{Weinberg:1977ma, Wilczek:1977pj},
\begin{equation}
m_a=6~\mu\mathrm{eV}\left(
\frac{f_a}{10^{12}\mathrm{GeV}}\right)^{-1},
\end{equation}
where $f_a$ is the axion decay constant. Couplings to other particles
are also suppressed by $f_a$, which is constrained from various
terrestrial experiments and astrophysical observations.
The most stringent lower bound $f_a\gtrsim 10^{10}$~GeV
comes from the duration time of neutrino burst from SN 1987A
\cite{Raffelt:1990yz,Turner:1989vc}. 

On the other hand, $f_a$ is also bounded from above based on
cosmological considerations. While couplings of axions to other particles
are very week, they can be copiously produced in high-energy
environments in the early Universe.  These axions contribute to the
energy density of dark matter. If inflation \cite{inflation}, 
which is essential in
current understanding of our Universe, takes place in the early epoch
and the PQ symmetry is broken during and after inflation, 
the axion field becomes
homogeneous over the observable Universe apart from quantum fluctuations, 
and undergoes coherent oscillation after the QCD phase transition 
with an initial value
$\mathcal O(f_a)$. Such coherently oscillating axion behaves as cold
dark matter (CDM), and should not dominate over the observed 
energy thereof.
From the cosmic microwave background (CMB) and other cosmological
observations, the abundance of the CDM is precisely constrained, which
gives a current bound on the decay constant $f_a$ as $f_a\lesssim
10^{11}$~GeV \cite{Wantz:2009it}. If the PQ symmetry is not restored
after the inflation, the isocurvature fluctuation of axion leads to
significant CMB anisotropies.  The amplitude of axion
isocurvature perturbation is proportional to the Hubble parameter in the
inflationary epoch $H_\mathrm{inf}$, and recent observations give a
constraint $H_\mathrm{inf}\lesssim10^{7}$~GeV for 
$f_a\simeq10^{10-12}$~GeV and misalignment angle 
$\theta_a \sim \mathcal O(1)$~\cite{Beltran:2006sq,Kawasaki:2007mb,Kawasaki:2008sn,Hikage:2008sk}.
Realizing such an inflation
model is generally difficult and requires  fine
tuning of parameters.  This calls up another possibility, where the PQ
symmetry is restored during or after the inflation. 
This happens rather naturally in the supersymmetric inflation models 
where the effective mass of $\mathcal O(H_{\rm inf})$ is induced for the PQ scalar
through supergravity effect and (if it is positive) restores the PQ
symmetry. 
Other models to restore the  symmetry during inflation have been proposed
in~\cite{Yokoyama:1988zza,Kawasaki:2010gv}. 

Since the PQ symmetry is a global U(1) symmetry, linear topological
objects called axionic strings can form when it breaks spontaneously.
When the spontaneous breaking of U(1)$_\mathrm{PQ}$ occurs in the
Universe, a cosmological network of axionic strings is formed. The axionic
string is a global string and an extended object with physical width
$d_\mathrm{string}\simeq1/\sqrt{2}f_a$.  While local strings can be
accurately simulated by using the Nambu-Goto action, it is difficult to
simulate the dynamics of global strings using string-based actions
\footnote{While global strings can be modeled by the Kalb-Ramond action,
it does not fit for numerical analysis.} and evolution of the string
network has been poorly understood.  A crucial feature that makes global
strings different from local strings is that a long-range force mediated
by massless NG boson operates between two global strings.  

In general, a scaling solution is characterized by the scaling 
parameter $\xi$, which we define as
\begin{equation}
\xi\equiv \frac{\rho_\mathrm{string}}{\mu_\mathrm{string}}t^2.
\label{eq:defscaling}
\end{equation}
$\xi$ is nothing but the average number of infinite strings in a Hubble
volume (strictly speaking, in a box with a volume $t^3$) and should be a
constant  of order of unity as long as a scaling solution is
realized. For local strings we find $\xi \sim 13$ from the 
simulations based on the Nambu-Goto action~\cite{NG} and $\xi \sim 6$ 
from  field-theoretic
simulations~\cite{Moore:2001px} in the radiation dominated
universe.\footnote{The reason of such a discrepancy is discussed and
hinted in Ref. \cite{Hindmarsh:2008dw}.} Whereas $\xi$ can be much smaller for
global strings
\cite{Yamaguchi:1998gx,Yamaguchi:1999yp,Yamaguchi:2002zv}.

In addition, the energy stored in global strings is released in a
different way from local strings. NG bosons dominantly carry away the
energy in global strings, while gravitational waves play the dominant
role for local strings. There has been a controversy about the energy
spectrum of axions radiated from axionic strings. Davis, Shellard and
co-workers insist that the spectrum has a sharp peak at the horizon
scale \cite{Davis:1989nj,Dabholkar:1989ju}.  On the other hand, Sikivie
and co-workers claim that it is proportional to the inverse momentum
\cite{Harari:1987ht,Hagmann:1990mj}. After the QCD phase transition,
radiated axions finally become CDM and their energy density is
proportional to the number density. The energy spectrum of axions is of
particular importance since the number density of radiated axions is
determined by the spectrum.

A solution for the controversy was given by Yamaguchi, Kawasaki and
Yokoyama \cite{Yamaguchi:1998gx} (hereafter YKY99). In their analysis, a
field theoretic simulation of axionic string was performed, which is a
first-principles calculation and least contaminated by theoretical
uncertainties.  Their analysis showed that the energy spectrum of axions
from strings are sharply peaked at a momentum around inverse of the horizon
scale and suppressed exponentially at higher momenta.  Therefore the
observed shape of the spectrum is in good agreement with the insistence
by Davis, Shellard and co-workers. In addition, YKY99 showed that a
scaling solution is also realized for global strings and find a much
smaller scaling parameter $\xi\simeq1.00\pm0.08$ compared with local
strings.\footnote{In more refined simulations, a slightly lower value of
$\xi$ has been obtained \cite{Yamaguchi:2002zv}.} This illuminates the
quantitative difference in dynamics of global strings from local
strings.

Our primary purpose of this paper is to update the analysis of YKY99.
While we perform a field theoretic simulation of axionic strings of
largest scale so far, we also develop several new techniques to improve
the accuracy of analysis.  One is a new method for
identification of strings in a simulation box. For field theoretic
simulation, this is a non-trivial task since field values are known only
at discrete spatial points. By checking the consistency with the
previous identification methods, reliability of our understanding of
string dynamics would be much improved.  In addition, we introduce a
pseudo power spectrum estimator (PPSE)~\cite{Wandelt:2000av,Hivon:2001jp} 
to remove the contamination of
string cores in estimation of the energy spectrum of radiated
axions. Since strings are highly-energetic objects, reliable estimate of
the energy density of free axions is difficult near string cores.
Therefore we should remove the regions near strings in estimation of the
energy spectrum. In YKY99, this was done by dividing the simulation box
into eight sub-boxes with the same size and estimating the spectrum using
only selected sub-boxes with no strings. In order to take
a larger simulation box and increase the statistics, we need a more
effective way to remove string cores.  This can be done by adopting
PPSE, which is often used in power spectrum estimation in CMB data
analysis to remove observed regions contaminated by foregrounds
including galactic emission and point sources~\cite{Hinshaw:2003ex}.

The organization of the paper is as follows. In Section \ref{sec:model}, 
we first give details of our model and the setup adopted in our field 
theoretic simulation. Then we describe our analysis method in Section
\ref{sec:analysis}. Particularly, we focus on methods adopted in
identification of strings and estimation of the energy spectrum.
Results of our analysis are presented in Section \ref{sec:results},
where  we mainly discuss the scaling property of axionic strings and 
the energy spectrum of radiated axions from
strings. Comparison with the previous result of YKY99 is also discussed here.
A constraint on $f_a$ and implications for the early Universe are 
discussed in Section \ref{sec:constraint}. 
The final section is devoted to summary and discussions.

In this paper, we assume flat Friedmann-Robertson-Walker Universe, with
a metric
\begin{equation}
ds^2=-dt^2+R(t)^2\delta_{ij}dx^idx^j,
\end{equation}
where $R(t)$ is the scale factor.
We also use a conformal time $\tau=\int^t_0 dt^\prime/R(t^\prime)$. A dot represents
derivative respect to the proper time, {\it i.e.} $\dot{}=\partial /\partial t$.

\section{Models and setup of field theoretic simulation}\label{sec:model}
We simulate dynamics of a complex PQ scalar field $\Phi(\vec x,t)$ 
with a Lagrangian density 
\begin{equation}
  \mathcal L=
  |\nabla_\mu\Phi|^2-V_\mathrm{eff}[\Phi;T], 
\end{equation}
with an effective potential at finite temperature $T$
\begin{equation}
  V_\mathrm{eff}[\Phi;T]=
  \frac{\lambda}{2}(|\Phi|^2-\eta^2)+\frac{\lambda}{3}T^2|\Phi|^2,
\end{equation}
where $\eta=f_a$ is the energy scale of PQ symmetry\footnote{
In general, $f_a=\eta/N_\mathrm{DW}$ where $N_\mathrm{DW}$ is the
number of degenerate vacua after QCD phase transition. 
Throughout this paper, we assume $N_\mathrm{DW}=1$, 
so that axionic domain walls quickly disappear after the QCD
phase transition (otherwise, it is cosmologically disastrous).
} and $\lambda$ is the self-coupling constant. The same potential is 
also adopted in YKY99. When the temperature is high enough 
$T>T_\mathrm{crit}\equiv\sqrt 3\eta$, 
$V_\mathrm{eff}$ has a minimum at $\Phi=0$ and 
U(1)$_\mathrm{PQ}$ is restored. At lower temperature $T<T_\mathrm{crit}$,
symmetry breaking minima appear at 
$|\Phi|=\eta\sqrt{1-(T/T_\mathrm{crit})^2}$.
The phase transition is of second order.
While we consistently adopt a particular set of model parameters 
$\eta=1.22\times 10^{16}$~GeV, $\lambda=1$ in our simulation, 
different choices of these parameters do not lead to any 
qualitative differences 
in physical consequences.

We simulate the evolution of $\Phi(\vec x,t)$ from the initial time 
$t_\mathrm{ini}=0.25t_\mathrm{crit}$ to the end time 
$t_\mathrm{end}=25t_\mathrm{crit}$,  denoting the 
time of the PQ phase transition by $t_\mathrm{crit}$, {\it i.e.}
$T(t_\mathrm{crit})=T_\mathrm{crit}$.
At the initial time $t_\mathrm{ini}$, 
we generate an initial condition for $\Phi(\vec x,t)$ and $\dot\Phi(\vec x,t)$.
At high temperature $T\gtrsim T_\mathrm{crit}$, $\Phi$ is approximately in thermal equilibrium,
so that $\Phi$ and $\dot\Phi$ can be regarded as Gaussian random variables.
By decomposing $\Phi$ into the real and imaginary parts, {\it i.e.} $\Phi=(\phi_1+i\phi_2)/\sqrt2$, 
correlation functions of $\phi_1$ and $\phi_2$ are given by 
\begin{eqnarray}
  \langle \phi_a(\vec x,t)^*
  \phi_b(\vec x^\prime,t)\rangle 
  &=&\delta_{ab}\int \frac{d^3k}{(2\pi)^3}
  \mathrm{e}^{-i\vec k\cdot(\vec x-\vec x^\prime)}
  \frac{1}{\omega(\vec k,t)}
  \frac{1}{\mathrm{e}^{\omega(\vec k,t)/T(t)}-1}, 
  \label{eq:initphi}\\
  \langle \dot\phi_a(\vec x,t)^*
  \dot\phi_b(\vec x^\prime,t)\rangle 
  &=&\delta_{ab}\int \frac{d^3k}{(2\pi)^3}
  \mathrm{e}^{-i\vec k\cdot(\vec x-\vec x^\prime)}
  \frac{\omega(\vec k,t)}{\mathrm{e}^{\omega(\vec k,t)/T(t)}-1},
  \label{eq:initphidot} \\
  \langle \phi_a(\vec x,t)\dot\phi_b(\vec x^\prime,t)\rangle&=&0, 
  \label{eq:initphicross}
\end{eqnarray}
where $\omega(\vec k,t)=\sqrt{k^2/R(t)^2+m(t)^2}$ is the energy of 
a Fourier mode with wave number $\vec k$, with the mass of $\Phi$ 
being denoted as $m(t)$, {\it i.e.} $m(t)^2=\partial^2 V_\mathrm{eff}/
\partial \Phi^*\partial\Phi|_{\Phi=0}=\lambda(T(t)^2/3-\eta^2)$. 
The subscripts $a$ and $b$ can be either $1$ or $2$.
We also note that in Eqs.~(\ref{eq:initphi})-(\ref{eq:initphicross}), 
we omit the contribution of vacuum fluctuations, which would 
not affect the classical dynamics of $\Phi$. 

The equation of motion for $\Phi$ is given by
\begin{equation}
  \left[\frac{\partial^2}{\partial t^2}
  +3H(t)\frac{\partial}{\partial t}-\frac{1}{R(t)^2}\nabla^2\right]
  \Phi(\vec x,t)=\frac{\partial V_\mathrm{eff}}{\partial\Phi^*}, 
  \label{eq:eom}
\end{equation}
where $H(t)=\dot R(t)/R(t)$ is the Hubble rate.
In our simulation, we integrate Eq.~\eqref{eq:eom} 
using the second order leapfrog scheme with a constant 
conformal time-step $\Delta \tau=2\times 10^{-3}\tau_\mathrm{crit}$.
The background Universe is dominated by radiation. There, the
Friedmann equation is given by
\begin{equation}
H(t)^2=\frac{8\pi G}{3}\frac{\pi^2}{30}g_*T^4,
\end{equation}
where $g_*$ is the number of relativistic degrees of freedom (d.o.f.).
In our simulation, we take it constant, $g_*=1000$, following YKY99.\footnote{%
While this value is
much larger than that predicted in Standard Model (and most of its
extensions), it does not affect our main results
thanks to the scaling properties of axionic strings.
}

Our lattice simulation has the number of grids 
$N_\mathrm{grid}=512^3$, which is larger than any other previous simulations.
We impose a periodic boundary condition and the
physical size of our simulation box at the end time $t_\mathrm{end}$
is taken to
$2(\tau_\mathrm{end}-\tau_\mathrm{crit})/\tau_\mathrm{end}=1.6$ 
times 
the horizon scale $1/H(t_\mathrm{end})$. 
With this choice of the box size, 
massless axions, 
which begin to be emitted at $t_\mathrm{crit}$, travel
for a distance less than half the size of our simulation box 
until $t_\mathrm{end}$. 
Therefore, we can avoid axions from overlapping around the box and 
causing unphysical boundary effects via the long range force 
mediated by them.
In this setup, the physical size of lattice spacing at the end time 
$t_\mathrm{end}$ turns out to be $R(t_\mathrm{end})\Delta x=
1.4d_\mathrm{string}$. 

Finally we comment on the  dynamical range of our simulation.
On one hand, our simulation box should be larger than the horizon, whose
comoving size scales as $1/R(t)H(t)\propto R(t)$, in order to avoid
boundary effects. On the other hand, the lattice spacing should be 
sufficiently smaller 
than the string width, whose comoving size scales as $1/R(t)$; otherwise,
strings cannot be resolved
 in the simulation box. As $R(t)$ becomes large, it turns out to be 
impossible to satisfy both of these two demands due to the finiteness of 
$N_\mathrm{grid}$. These determine the dynamical range.

\section{Analysis method}\label{sec:analysis}
Our field-theoretic simulation is a first-principles 
calculation and  free from theoretical uncertainties.
Several difficulties, however, arise in extracting the
physically relevant quantities from the given data
of $\Phi$ and $\dot\Phi$ at discrete points.
As discussed in Introduction, identification
of strings and estimation of energy spectrum of 
axions are of primary importance. 
So we describe the details of these matters in this section.

\subsection{Identification of strings}\label{sec:id}
Several methods for string
identification have been developed so far. In
Refs. \cite{Yamaguchi:1998gx,Yamaguchi:1999yp}, a lattice was identified
as a part of a string based on the value of the potential energy
there. Counting number of lattices thus identified as containing 
strings, they estimated the scaling parameter. Although the overall
features were traced reasonably well, this method had some problems in
that string segments were 
occasionally found disconnected and it might
overestimate the scaling parameter because the number of
the lattices penetrated by a string was used rather than the length of
the string itself. 

Later, two of the present authors (MY and JY) developed a
new method of string identification \cite{Yamaguchi:2002zv}, which uses
the phase of the fields. In this method, first  a quadrate, which
a string penetrates, is identified
by dividing the phase into three zones with unequal spans\footnote{If
one divided the phase into three equal zones, one would 
occasionally identify a quadrate as
containing a string even if either $\phi_1$ or $\phi_2$ takes the same
sign at its four corners. So the authors of \cite{Yamaguchi:2002zv} had
to divide the phase into uneven zones.  Nevertheless it did not cause
any artifacts in the final results.} and
monitoring the phase rotation just as in the Vachaspati-Vilenkin
algorithm \cite{Vachaspati:1984dz}. 
Then, the position of the string in each
quadrate is determined by the zeros of the two real fields,
$\phi_1$ and $\phi_2$. 
These complicated identification scheme guarantees the connectedness of
strings and makes it possible to calculate
 the string velocity, the intercommutation rate, and the NG
boson emission rate as well as  the scaling
parameter more precisely. 
Thereby it enables us to analyze the evolution of 
the string network from the Lagrangian viewpoint to clarify its
fundamental
characters.
\if
However, we did not assign an equal range $2\pi/3$ of the relative phase
to all zones because we would occasionally identify a quadrate as
containing a string even if either $\phi_1$ or $\phi_2$ takes the same
sign at its four corners. Such a biased (uneven) assignment of the
relative phase may cause some artificial effects. 
Fortunately, we found no significant differences, although we tried
another division of the relative phase.  The complete connection of the
strings are indispensable for the purpose of
Ref. \cite{Yamaguchi:2002zv}, that is, to clarify the quantitative
nature of the cosmological evolution of a global string network from a
Lagrangian viewpoint. However, in this paper, we have only to find a
quadrate that a string penetrates and to estimate an accurate value of
the scaling parameter.
\fi

For our practical purposes here, however, we do not need to adopt
such a complicated and time-consuming procedure albeit its accuracy,
because the final error is dominated by the finiteness of the
simulation box rather than the identification scheme we use.  
We therefore adopt a new and much more efficient method in which
we do not need to divide the phase into fixed uneven zones.
\if
Therefore, we introduce a new method that is appropriate 
for the present purpose.  \Red{This method circumvents the
biased (uneven) assignment of the relative phase and can reduce the
computation time.} \remb{What does `relax' means?} 
\fi

\begin{figure}[tb]
\begin{center}
  \begin{tabular}{ccc}
    real space & \hspace{1cm} & field space \\
    \scalebox{0.5}{\includegraphics{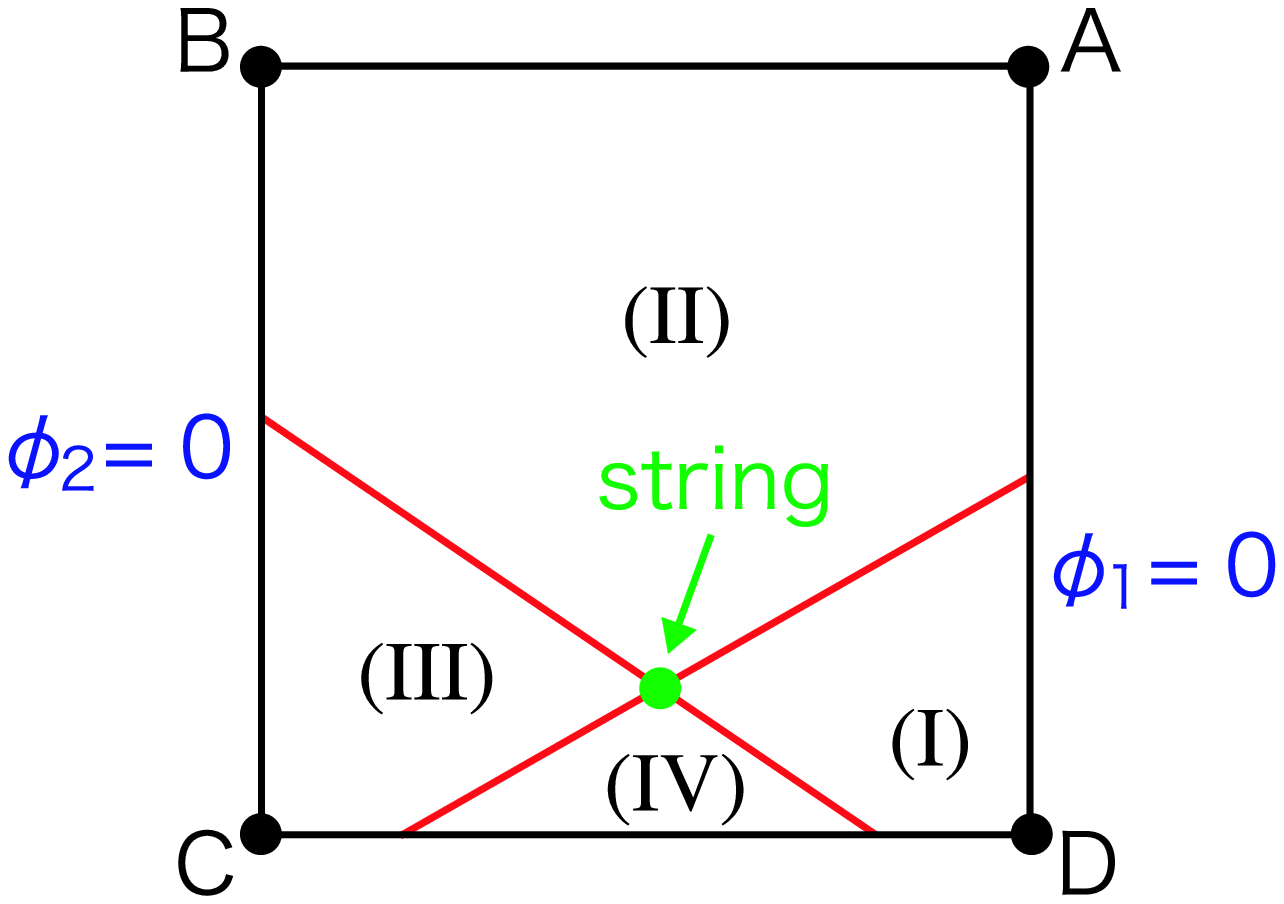}} & &
    \scalebox{0.5}{\includegraphics{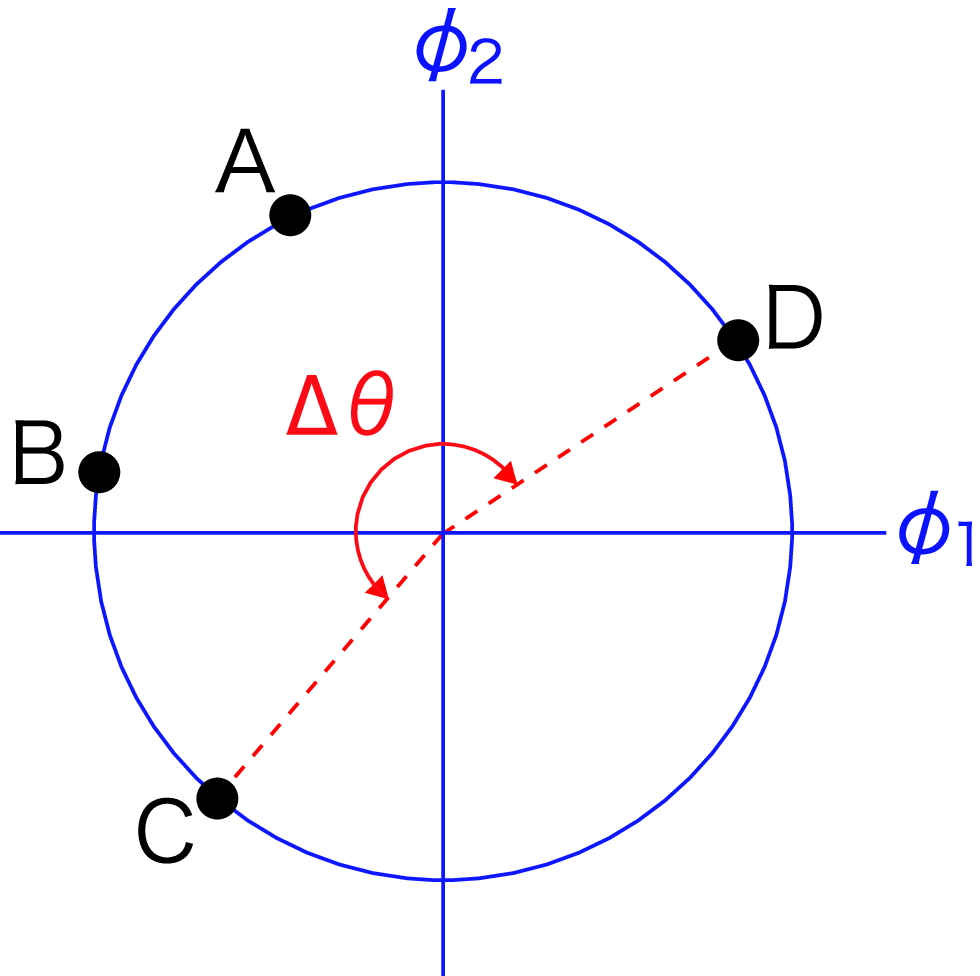}}
  \end{tabular}
  \caption{Schematic view of our method for string identification.
  (Left) Shown is a quadrate in real space penetrated by a 
  string (green point). The loci of $\phi_1=0$ and 
  $\phi_2=0$ (red lines) intersect with each other at the string. 
  Note that the string does not necessarily penetrate the quadrate 
  perpendicularly. When the phase of $\Phi$ continuously changes 
  around the string isotropically, phases at the opposite sides of an arbitrary 
  line on the quadrate intersecting the penetration point differ
  by $\pi$. Greek numbers (I), (II), (III) and (IV) in the left panel 
  shows the quadrants of $(\phi_1,\phi_2)$ in these regions.
  (Right) Shown is the mapping of four vertices in the 
  field space. The minimal phase range containing the images of 
  these four vertices is indicated with a red arrow. By observing
  whether the phase difference $\Delta\theta$ in this minimal range 
  is larger than $\pi$, we identify quadrates penetrated by strings.
  }
\label{fig:identify}
\end{center}
\end{figure}

Let us start by considering a small quadrate whose
four vertices are the neighboring grids in the simulation box (See
the left panel of Figure \ref{fig:identify}).  
Since the field values $\Phi(\vec x,t)$ at 
these vertices (A, B, C and D in the figure) are known from simulation, 
we can map these vertices into the field space 
(Right panel of the figure). While there are several ways to
take a region of phase which contains the images of these four vertices,
we denote the range of minimal one by $\Delta\theta$, which is explicitly
shown in the right panel of Figure \ref{fig:identify}.  
As shown in the figure,
if a string penetrates the quadrate,
$\Delta\theta$ should be larger than $\pi$. This can be easily
understood by drawing a line on the quadrate running through the
penetration point of the string. When the quadrate is small enough for
the continuous change of the phase to be regarded as isotropic around
the string, the phases at opposite sides of an arbitrary line
intersecting the penetration point (for example, lines of $\phi_1=0$ and
$\phi_2=0$ shown in the left panel of Figure \ref{fig:identify}) 
differ by $\pi$.  Whatever line we draw, each region of the
quadrate divided by the line contains at least one vertex.  Therefore,
the minimal phase difference cannot be smaller than $\pi$.  
As long
as the continuous phase change around strings is isotropic, the opposite
is also true, {\it i.e.} if $\Delta\theta>\pi$, 
the quadrate is penetrated by
a string. Therefore we can use $\Delta\theta$ as a criterion for
identification of strings.  One interesting point is that our method is
also applicable for any convex polygons other than quadrate. In
addition, our method is invariant under the global rotation of the phase
of $\Phi$, which is a desirable feature in identification of global
strings.

Of course there are loopholes in our new method. Our method assumes the
isotropy of the continuous phase change around strings.  This is valid
when a string is nearly straight and there are no other strings
nearby. However, this assumption breaks down around regions where
strings are colliding with others and/or strings have small curvature
radii. In rare cases, two or more strings can penetrate one single
quadrate in such regions. Therefore, some strings are not connected
completely in this method. However, fortunately, the fraction of such
regions is quite small; it is at most 1\% when the system of strings
relaxes into the scaling regime. Moreover, at around these regions,
phase of $\Phi$ changes very violently if not isotropically. Therefore
the criterion $\Delta\theta>\pi$ is still useful to identify strings,
although it to some extent over- or underestimate penetration of
strings. In the end, we can safely say that the fractions of both
omission and commission are less than $1$\% in our method.

When a quadrate is expected to be penetrated by strings, then we
determine penetration points of strings in it.  Our
determination method of the penetration points is basically the same as
in \cite{Yamaguchi:2002zv}: we first compute the points with $\phi_1=0$
or $\phi_2=0$ on the boundaries of a quadrate using linear interpolation
of $\Phi$; then we identify the intersection of lines that connect two
points with either $\phi_1=0$ or $\phi_2=0$ as the penetration point
(See the left panel of Figure \ref{fig:identify}).  It sometimes happens
that two or more strings are penetrating a single quadrate. In such
cases, there are four (not two) points with either $\phi_1=0$ or
$\phi_2=0$ on the boundaries of the quadrate.  We compute the sign of
$\phi_1$ or $\phi_2$ at the center of the quadrate by averaging over
those at four vertices, and then determine two pairs of points each of
which we connect so that the sign of $\phi_1$ or $\phi_2$ at the center
is respected. Sometimes intersections of $\phi_1=0$ and $\phi_2=0$ are
found lying outside the quadrates, but more than 99 \% of penetration
points are found lying inside each quadrate. In conjunction with the
omission and commission rates in identification of penetrated quadrates,
we estimate our method can determine the positions of strings with at least
99 \% accuracy. This is highly sufficient to analyze the dynamics of
strings and estimate the energy spectrum of radiated axions with masking
of strings.

\subsection{Estimation of the energy spectrum of radiated axions}
\label{sec:estimation}

\subsubsection*{Energy spectrum of axions}
\label{sec:spectrum}

From the simulated data of $\Phi$ and $\dot\Phi$, we obtain 
the time derivative of the axion field
\begin{equation}
  \dot a(\vec x,t)=
  \mathrm{Im}\left[\frac{\dot\Phi}{\Phi}(\vec x,t)\right], 
\end{equation}
and its Fourier component is given by
\begin{equation}
  \dot a(\vec k,t)=
  \int d^3x\mathrm{e}^{i\vec k\cdot\vec x}\dot a(\vec x,t),
\end{equation}
where $\vec k$ is a comoving wave number.

Assuming the statistical isotropy and homogeneity, 
the two-point correlation function in Fourier space
can be represented in terms of the power spectrum $P(k)$,  
\begin{equation}
  \frac{1}{2}\langle \dot a(\vec k,t)^*\dot a(\vec k^\prime, t)\rangle=
  \frac{(2\pi)^3}{k^2}\delta^{(3)}(\vec k-\vec k^\prime)P(k,t), 
  \label{eq:power}
\end{equation}
where the brackets $\langle \cdot\rangle$ represent an ensemble average, 
{\it i.e.} an average over infinite realizations.
While homogeneity and isotropy is in reality broken by the finiteness of the simulation volume,
they are approximately valid as long as the wave number $k$ is not very small ($k\lesssim1/L$)
nor very large ($k\gtrsim1/\Delta x$).
The mean kinetic energy of the axion $\bar \rho(t)$ 
is nothing but the ensemble average of energy density $\rho(\vec x,t)=\frac{1}{2}\dot a(\vec x,t)^2$, 
and can be rewritten in terms of $P(k)$, 
\begin{eqnarray}
  \bar \rho(t)&=&
  \left\langle \frac{1}{2}\dot a(\vec x,t)^2\right\rangle\notag\\ 
  &=&\int \frac{d^3k}{(2\pi)^3}
  \int \frac{d^3k^\prime}{(2\pi)^3}
  \langle\dot a(\vec k,t)^*\dot a(\vec k^\prime,t)
  \rangle
  \mathrm{e}^{i(\vec k^\prime-\vec k)\cdot\vec x}\notag\\
  &=&\int\frac{dk}{2\pi^2}P(k,t).
\end{eqnarray}
This shows that the power spectrum $P(k)$ defined in Eq.~\eqref{eq:power} 
is nothing but the energy spectrum of axion field.

\subsubsection*{PPSE of the energy spectrum of radiated axions}
\label{sec:ppse}

What we concern is the energy spectrum  $P_\mathrm{free}(k)$ of free 
axions radiated from strings.
However, $\dot a$ is also induced by moving strings. Separation of 
contribution from free axions and contamination from string motion is 
quite difficult. Moreover, the energy density associated with 
moving strings is so large that it can  
easily dominate over that.
Therefore it is essential to remove regions around strings 
in estimation of $P_\mathrm{free}(k)$.
In the present work we adopt PPSE to deal with this issue.

Let us start from the time derivative of the axion field 
$\dot a(\vec x,t)$ at a fixed time $t$. 
At most points in the simulation box, $\dot a$ represents the time
derivative of free axions, which we denote by $\dot a_\mathrm{free}$.
However, near a string, $\dot a$ is also
induced by motion of strings, so that 
\begin{equation}
  \dot a(\vec x,t)=
  \dot a_\mathrm{free}(\vec x,t)+(\mbox{contamination from strings}).
\end{equation}
In the same way as in Eq. \eqref{eq:power},
$P_\mathrm{free}(k)$ is defined by
\begin{equation}
  \frac{1}{2}\langle \dot a_\mathrm{free}(\vec k,t)^*\dot 
  a_\mathrm{free}(\vec k^\prime, t)\rangle=
  \frac{(2\pi)^3}{k^2}\delta^{(3)}(\vec k-\vec k^\prime)
  P_\mathrm{free}(k,t). 
  \label{eq:power_free}
\end{equation}
For a while, we drop the argument $t$ because the energy spectrum
is calculated at each fixed time. 

Fortunately, the contamination is localized around
string cores. Therefore we can mask the string contamination 
by adopting a window function
\begin{equation}
  W(\vec x)=\begin{cases}
  0 & (\mbox{near strings}) \\
  1 & (\mbox{elsewhere}) 
  \end{cases}.
\end{equation}
If the mask covers regions large enough so that the string contamination is 
removed to a negligible level, masked $\dot a$ has contribution only from free axions.
Thus we obtain
\begin{equation}
  \tilde{\dot a}(\vec x)\equiv
  W(\vec x)
  \dot a(\vec x)=
  W(\vec x)
  \dot a_\mathrm{free}(\vec x).
\end{equation}
In the Fourier space, $\tilde{\dot a}(\vec k)$ is the convolution 
of $W(\vec k)$ and $\dot a_\mathrm{free}(\vec k)$, 
\begin{equation}
  \tilde{\dot a}(\vec k)=
  \int\frac{d^3k^\prime}{(2\pi)^3}W(\vec k-\vec k^\prime)
  \dot a(\vec k^\prime)=
  \int\frac{d^3k^\prime}{(2\pi)^3}W(\vec k-\vec k^\prime)
  \dot a_\mathrm{free}(\vec k^\prime).
\end{equation}

We can straightforwardly calculate the `masked' energy spectrum 
in a given simulation box, 
\begin{equation}
  \tilde P(k)\equiv \frac{k^2}{V}\int \frac{d\Omega_k}{4\pi}
  \frac{1}{2}\left| \tilde{\dot a}(\vec k)\right|^2, \label{eq:masked}
\end{equation}
where $V$ is the comoving volume of the simulation box and the
integration
is performed over the angular direction of $\vec k$ with the
solid-angle element $d\Omega_k$.
However, as shown in Appendix \ref{app:ppse}, the masked spectrum 
is biased, {\it i.e.} $\langle \tilde P(k)\rangle\ne P_\mathrm{free}(k)$. 
This is because the wave number $\vec k$ to which 
$\dot a_\mathrm{free}(\vec k^\prime)$ contributes
can be different from the original one $\vec k^\prime\ne \vec k$ due to 
the mode-mixing induced by the window function $W(\vec k-\vec k^\prime)$. 
Such a mode mixing can be corrected using PPSE. 
We define a PPSE of $P_\mathrm{free}(k)$, 
\begin{equation}
  \hat P(k)\equiv\frac{k^2}{V}\int \frac{dk^\prime}{2\pi^2}
  M^{-1}(k,k^\prime)\tilde P(k^\prime),
  \label{eq:ppse}
\end{equation}
where $M^{-1}(k,k^\prime)$ is defined so that it satisfies
\begin{equation}
  \int \frac{k^{\prime2}dk^\prime}{2\pi^2}
  M^{-1}(k,k^\prime)M(k^\prime,k^{\prime\prime})
  =\frac{2\pi^2}{k^2}\delta(k-k^{\prime\prime}), \label{eq:inverse}
\end{equation}
with 
\begin{equation}
  M(k,k^\prime)\equiv\frac{1}{V^2}
  \int \frac{d\Omega_k}{4\pi}\frac{d\Omega_{k^\prime}}{4\pi}
  \left| W(\vec k-\vec k^\prime)\right|^2. \label{eq:winmat}
\end{equation}
Then $\hat P(k)$ is unbiased,  {\it i.e.} 
$\langle \hat P(k)\rangle=P_\mathrm{free}(k)$. 
The proof is given in Appendix \ref{app:ppse}.

\subsubsection*{Pipeline and validity check}
\label{sec:pipeline}

\begin{figure}[!tb]
\begin{center}
  \scalebox{0.6}{\includegraphics{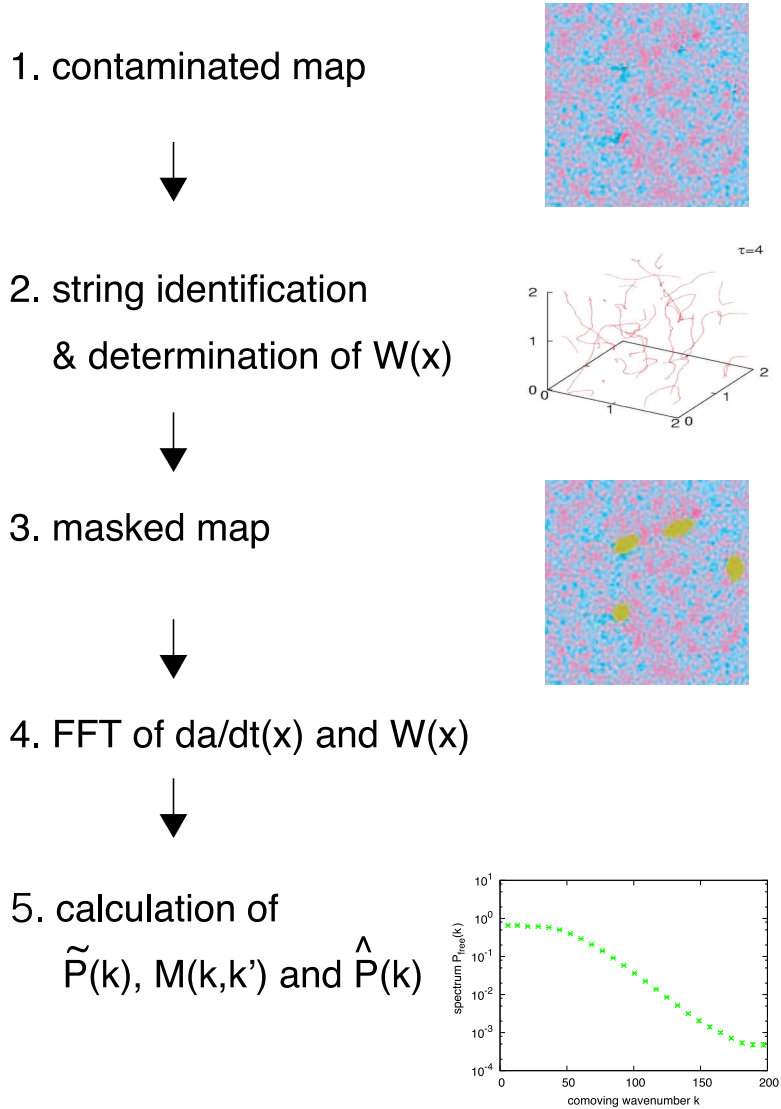}}
  \caption{Schematic overview of our pipeline.}
\label{fig:pipeline}
\end{center}
\end{figure}

Figure \ref{fig:pipeline} shows our pipeline. 
For each single realization of the field theoretic simulation, 
a configuration of $\dot a(\vec x)$ is given at first. 
Then we identify strings in the simulation box
and determine $W(\vec x)$. 
Practically, we mask grid points separated away 
from any strings by less than $3d_\mathrm{string}$.
By applying masking, we obtain $\tilde {\dot a}(\vec x)$.
After implementing the Fourier transformation of 
$\tilde {\dot a}(\vec x)$, we obtain
$\tilde P(k)$. We also implement the Fourier transformation of $W(\vec k)$, 
and calculate $M(k,k^\prime)$. By multiplying $\tilde P(k)$ by inverse of 
$M(k,k^\prime)$, we obtain $\hat P(k)$. 

\begin{figure}[tb]
\begin{center}
  \scalebox{1.5}{\includegraphics{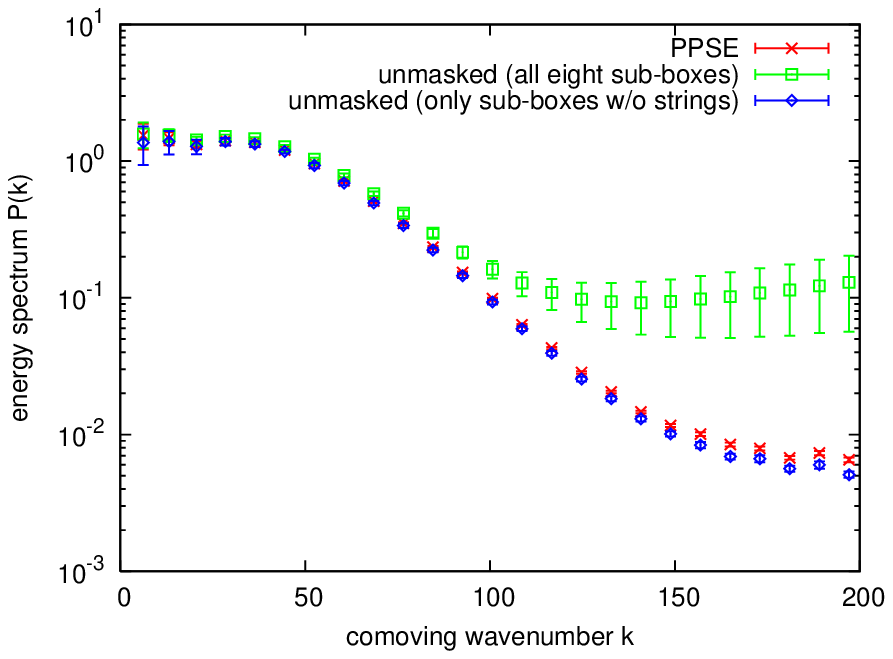}}
  \caption{Validity check of our estimation method using PPSE.
  Three different spectra are plotted (See text for details).
  Only statistical errors are shown; bars corresponds to 
  the square root of the diagonal components of covariance matrices.
  }
\label{fig:check}
\end{center}
\end{figure}

Now we check the validity of our method. 
For this purpose,
we performed a field theoretic simulation with lower resolution
than that described in Section~\ref{sec:model}. 
Here we take the lower number of grids ($N_\mathrm{grid}=256^3$),
decreased dynamical range $t_\mathrm{end}=16t_\mathrm{crit}$
and smaller box size, $1.0/H(t_\mathrm{end})$.
Other parameters are the same as in Section \ref{sec:model}.
In estimation of the spectrum, we first divide the whole simulation 
box into eight sub-boxes of the same size. Then in each sub-box, 
we calculated two different spectra estimated with or without masking.
We note that while the energy spectrum might be affected 
by boundary effects, 
this does not cause any problem; what we concern here is verification 
of the method, not the physical consequences.
In Figure \ref{fig:check}, we plot three different spectra estimated 
at $t=16t_\mathrm{crit}$ obtained by averaging over 
20 independent realizations. 
The spectrum in red crosses corresponds to our PPSE $\hat P(k)$ with masking.
The spectra in green boxes and blue diamonds correspond 
to energy spectra calculated without masking of strings.
While all sub-boxes are used in calculation of the spectrum in green boxes, 
only sub-boxes found without strings in them are selectively used 
for one with blue diamonds. Therefore 
the blue spectrum is exactly what is calculated in YKY99.
The comoving wave number $k$ and estimated energy 
spectra $P(k)$ are normalized in
$\tau^{-1}_\mathrm{crit}$ and $\tau_\mathrm{crit}^{-3}$, respectively.

First of all, we see excellent agreement between our PPSE $\hat P(k)$ (red crosses) 
and the spectrum obtained only from sub-boxes without strings (blue diamonds) in
Figure \ref{fig:check}. Therefore our estimation method of 
the energy spectrum
of free axion can be regarded as working quite well. 
The differences are less than 10 \% at $k\tau_\mathrm{crit}\le100$
and at most 25 \% at $k\tau_\mathrm{crit}\le 200$. We regard this discrepancy 
as systematic errors in estimation of $P_\mathrm{free}(k)$.
We also observe the energy spectrum estimated without masking (green boxes)
has significant contribution from string cores at large $k$. 
This is because the phase of PQ field at some point fixed 
in a comoving frame rapidly changes 
when a string passes nearby. 
However, such a string
does not radiate axions and free axions are not responsible for 
the observed large $\dot a$. Since such large $\dot a$ induced 
by motion of strings is concentrated around strings, 
it gives power at large $k$. Since the contribution of string cores is
also significant at moderate scales $k\tau_\mathrm{crit}\lesssim100$, it is essential 
to remove the contamination from string cores to estimate the spectrum
of energy released by axion emission, which we will discuss in Section 
\ref{sec:net}.

As a final remark of verification check, we note that at small scales $k\tau_\mathrm{crit}\gtrsim150$, 
even red and blue spectrum differ from the true spectrum which can be
calculated from whole the simulation box. This is because we 
performed discrete Fourier transformation in each sub-boxes assuming
a periodic boundary condition. Since the boundary condition does not 
hold actually within each sub-box, discontinuities in $\dot a(\vec x)$ 
arise at boundaries and this gives additional power to the energy 
spectrum at large $k$.\footnote{
This is the very reason that the energy spectrum shown in Figure 2 of YKY99
shows deviation from exponential behavior at large wave numbers, 
which is not mentioned in YKY99.}.
True energy spectra should have to some extent lower amplitudes at
large $k$ (See Figure \ref{fig:spec}).

\section{Results}\label{sec:results}

\subsection{Scaling property of axionic strings}\label{sec:dynamics}
\begin{figure}[!tb]
\begin{center}
  \scalebox{1.1}{\includegraphics{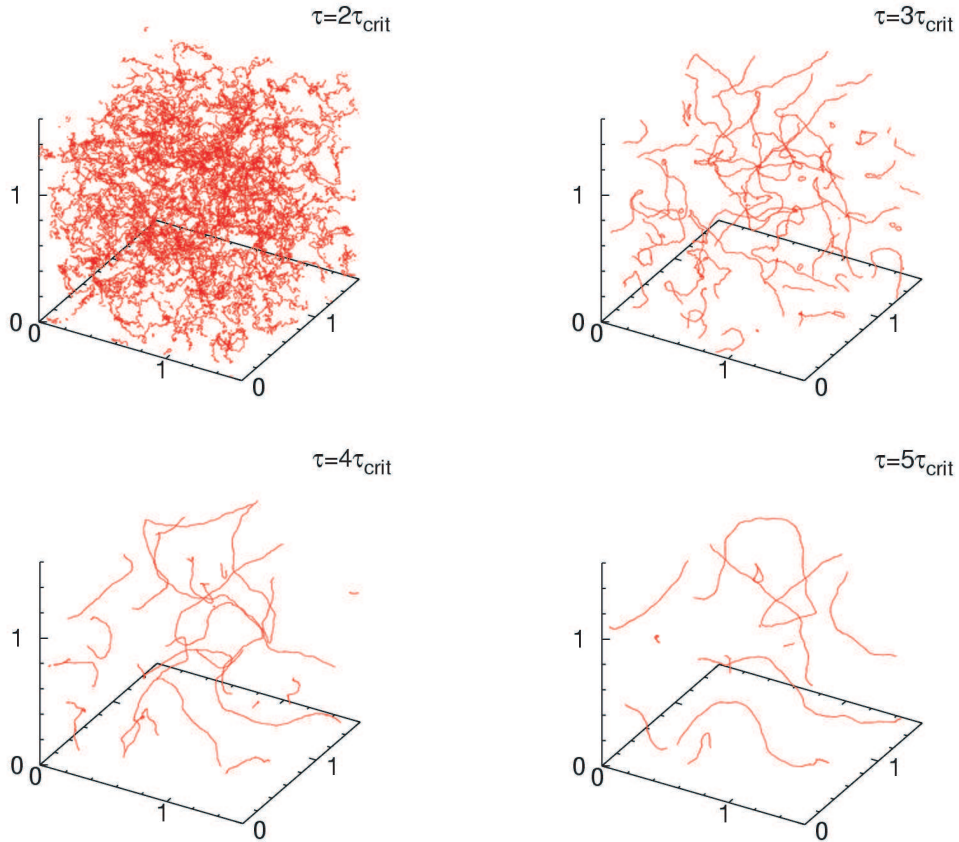}}
  \caption{Visualization of one realization from our field theoretic 
  simulation.
  Red line corresponds to axionic string identified by our method 
  discussed in Section \ref{sec:id}. $\tau$ in the top right 
  of each panel is the conformal time of each time slice, 
  which can be translated into the proper 
  time $t$ via a relation in radiation domination 
  $t/t_\mathrm{crit}=\tau^2/\tau_\mathrm{crit}^2$.  
  The spatial scale shows a comoving length
  in unit of the horizon size at $t_\mathrm{end}=25t_\mathrm{crit}$.
  }
\label{fig:slides}
\end{center}
\end{figure}

Figure \ref{fig:slides} is a visualization of one realization 
in our simulation. There, red points are the positions of axionic strings 
determined by the method explained in the previous section. 
The points clearly form line-shaped objects regarded as axionic strings, 
which shows good resolution in our simulation as well as excellent 
string identification.

\begin{figure}[tb]
\begin{center}
  \scalebox{1.5}{\includegraphics{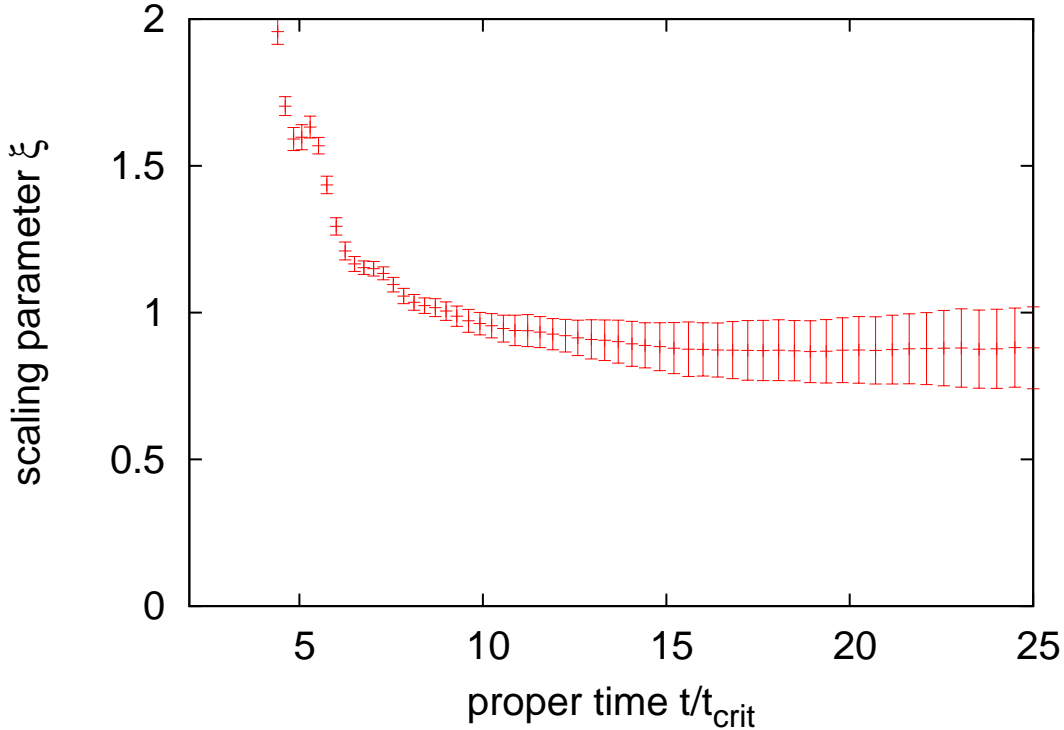}}
  \caption{Time evolution of the scaling parameter $\xi$ obtained
  by averaging over 20 realizations. Note that data points 
  are not homogeneously placed in $t$, but in $\tau\propto \sqrt t$.
  }
\label{fig:scaling}
\end{center}
\end{figure}

By linearly connecting the penetration points of strings in neighboring
quadrates, we can estimate the total length of strings in the simulation
box.  Recalling that $\rho_\mathrm{string}/\mu_\mathrm{string}$ in
Eq.\eqref{eq:defscaling} corresponds to the mean physical length of
strings in unit physical volume, we can compute the scaling parameter
$\xi$ from the total length of strings.  In Figure \ref{fig:scaling}, we
plotted the time evolution of $\xi$ which is obtained by averaging over
20 realizations with the setup explained in Section
\ref{sec:model}. We first see that $\xi$ stays constant for
$t\gtrsim 10t_\mathrm{crit}$. 
This shows that the system of axionic strings relaxes into 
the scaling regime.
At $t=25t_\mathrm{crit}$,
we obtain
\begin{equation}
   \xi=0.87\pm0.14.
   \label{eq:result_xi}
\end{equation}
Our result shows good agreement with \cite{Yamaguchi:2002zv}, 
which gives $\xi\simeq0.8$. While YKY99 gives
a slightly higher value $\xi=1.00\pm0.08$, the difference
is not significant at all. 

It might be curious that the error in $\xi$ grows as $t$ 
increases in Figure \ref{fig:scaling}.
This is due to a statistical reason; since the horizon scale becomes 
larger as time advances, the number of independent horizon volumes 
in the simulation box effectively becomes smaller at later times. 
Therefore, variation in $\xi$ among realizations becomes larger 
as $t$ increases, which
increases the errors in $\xi$ averaged over realizations.
Moreover, our estimate of $\xi$ in Eq. \eqref{eq:result_xi} 
can be regarded as rather conservative one.

\subsection{Net energy spectrum of radiated axions}\label{sec:net}

\begin{figure}[tb]
\begin{center}
  \begin{tabular}{cc}
    \scalebox{1}{\includegraphics{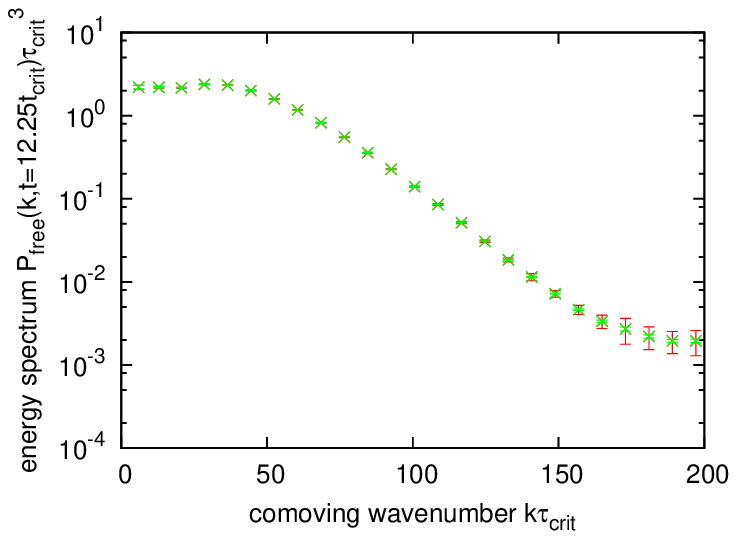}} &
    \scalebox{1}{\includegraphics{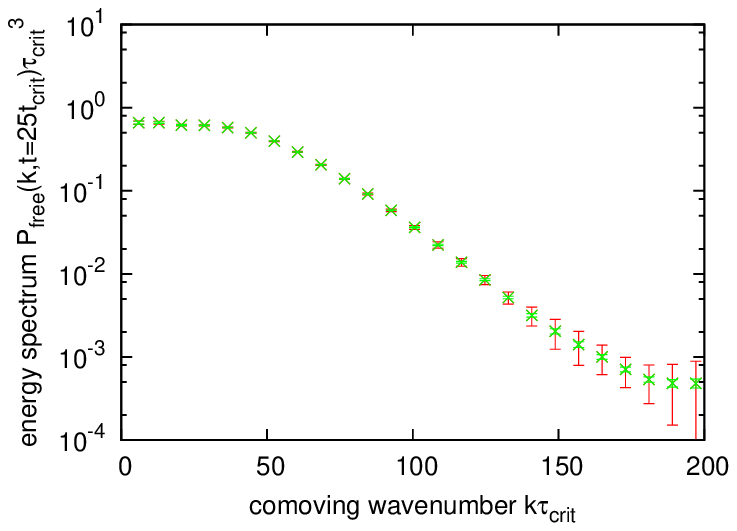}}
  \end{tabular}
  \caption{Energy spectra of radiated axions at 
  $t_1=12.25t_\mathrm{crit}$ (left)
  and $t_2=25t_\mathrm{crit}$ (right).   
  Green bars correspond to statistical errors alone; 
  red error bars also count systematic errors suggested in Section 
  \ref{sec:pipeline} as well as statistical ones.
  }
\label{fig:spec}
\end{center}
\end{figure}

Figure \ref{fig:spec} shows the energy spectra of radiated axions
at $t_1=12.25t_\mathrm{crit}$ (left) and $t_2=25t_\mathrm{crit}$ (right), 
that are estimated from the 20 realizations used in Section \ref{sec:dynamics}.
The amplitude of energy spectrum at $t_2$ is 
about $(t_1/t_2)^2\simeq0.24$ times that at $t_1$. This is because
the energy density of free axions  scales as $R^{-4}(t)$, without emission 
or absorption. We see a clear exponential behavior at large $k$ after the 
removal of the contamination from strings. 

As we see in Section \ref{sec:dynamics}, the system of axionic strings
are already in the scaling regime. Most of axions at this epoch are 
however emitted before the settlement into the regime. 
In order to extract the energy spectrum of axions radiated during 
the scaling regime, we need to differentiate the energy spectra at 
different times.
We define the differential spectrum of 
radiated axion between $t_1$ and $t_2$, 
\begin{equation}
\Delta P_\mathrm{free}(k;t_1,t_2) \equiv
R^4(t_2)P_\mathrm{free}(k,t_2)-R^4(t_1)P_\mathrm{free}(k,t_1).
\end{equation}
If there are no emission nor absorption of axions, 
the energy density of axion scales as $R^{-4}(t)$. Therefore, 
$\Delta P_\mathrm{free}(k,t)$ is the net energy spectrum of 
axions radiated from strings.

\begin{figure}[tb]
\begin{center}
  \scalebox{1.5}{\includegraphics{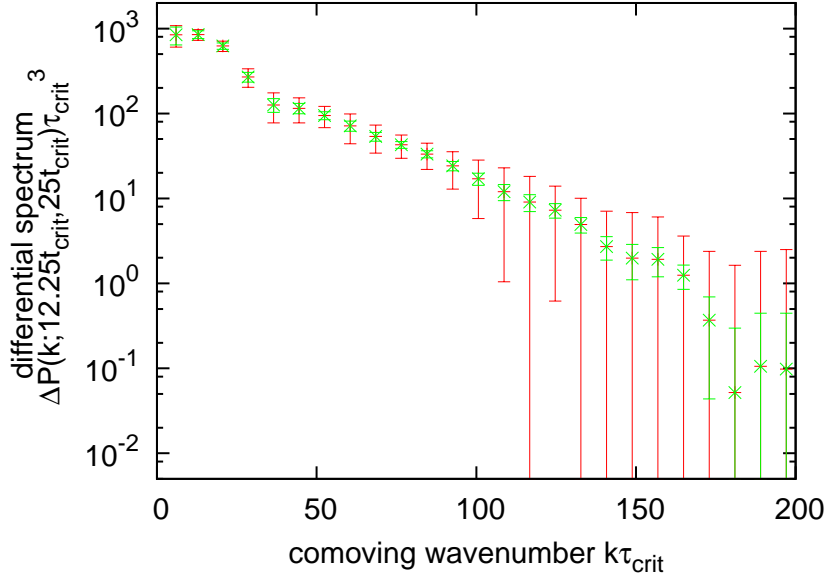}}
  \caption{Differential energy spectrum of radiated axions between 
  $t_1=12.25t_\mathrm{crit}$ and $t_2=25t_\mathrm{crit}$. 
  Errors shown are estimated from the quadrature of those 
  shown in Figure \ref{fig:spec}. Green (red) bars correspond
  to statistical errors alone (statistical and systematic errors).
  Note that the scale in $y$-axis is arbitrary 
  since the scale factor $R(t_\mathrm{crit})/R_0$ is not well-defined.}
\label{fig:diffspec}
\end{center}
\end{figure}

In Figure \ref{fig:diffspec}, we plotted the differential spectrum 
between $t_1$ and $t_2$. 
We observe that the net energy spectrum is 
sharply peaked at around the horizon,
which corresponds to $k=3.6\tau^{-1}_\mathrm{crit}$ at $t=t_1$
($k=2.5\tau^{-1}_\mathrm{crit}$ at $t=t_2$), 
and its amplitude 
is exponentially suppressed toward higher $k$. This is consistent with 
YKY99.

To calculate the number density of axions created from 
strings, the important quantity is
the mean reciprocal comoving momentum of radiated axion defined by
\begin{equation}
  \overline{k^{-1}(t)}\equiv
  \frac
  {\int \frac{dk}{2\pi^2}\frac{1}{k}\Delta P_\mathrm{free}(k,t)}
{\int \frac{dk}{2\pi^2}\Delta P_\mathrm{free}(k,t)}.
\end{equation}
By fitting the ratio of $R(t)\overline{k^{-1}(t)}$ to the 
horizon scale $t/2\pi$, we obtain
\begin{equation}
  \frac{R(t)\overline{k^{-1}(t)}}{t/2\pi}\equiv\epsilon^{-1}=0.23\pm0.02.
  \label{eq:result_eps}
\end{equation}
This is consistent with
$\epsilon^{-1}=0.25\pm0.18$ obtained by YKY99.
However, the uncertainty is reduced by an order of 
magnitude owing to the increase in statistics, which
is brought about by use of PPSE.

\section{Constraint on the axion decay constant}\label{sec:constraint}
Using the result of the previous section, we can derive a 
constraint on $f_a$. After the PQ phase transition, the system of 
axionic strings reaches the
 scaling regime and continuously emit massless axions.
After the QCD phase transition, the axionic string-wall system is generated.
After $t_w$, when the tension of axionic domain walls $\sigma_\mathrm{wall}$
starts to dominate the system, {\it i.e.} $\sigma_\mathrm{wall}(t_w)=
\mu_\mathrm{string}(t_w)/t_w$, 
the wall-string system quickly disappears. 
Emission of axions from strings continues until $t_w$, and 
the largest fraction of the axions are produced just before $t_w$.
After $t_w$, axions becomes non-relativistic due to the finite mass.
Today, axions exist as CDM in the Universe.

In Appendix \ref{app:energy}, we give a detailed derivation of the
density parameter of CDM axions radiated from axionic strings, 
adopting the axion mass at finite temperature from recent studies
\cite{Wantz:2009mi,Wantz:2009it}. 
By substituting Eqs. \eqref{eq:result_xi} and \eqref{eq:result_eps}
obtained from our simulation into Eq. \eqref{eq:omega_axion}, 
we obtain
\begin{equation}
  \Omega_\mathrm{axion}h^2=(1.66\pm0.25)
  \gamma
  \left(\frac{g_{*w}}{70}\right)^{-0.31}
  \left(\frac{\Lambda}{400\mathrm{MeV}}\right)
  \left(\frac{f_a}{10^{12}\mathrm{GeV}}\right)^{1.19},
  \label{eq:result_omega}
\end{equation}
where $\gamma$, $g_{*w}$ and $\Lambda$ 
are the dilution factor, the number of relativistic d.o.f. at 
$t_w$, and the scale of the QCD phase transition, 
respectively (See Appendix \ref{app:energy}).

$\Omega_\mathrm{axion}h^2$ should be smaller than
the observed $\Omega_\mathrm{CDM}h^2$. 
Recent cosmological observations \cite{Komatsu:2010fb} 
give $\Omega_\mathrm{CDM}h^2=0.11$ 
(assuming flat power-law $\Lambda$CDM model). Therefore 
we can translate Eq. \eqref{eq:result_omega} into constraint on $f_a$.
Assuming no entropy dilution ($\gamma=1$), 
we obtain $f_a\le1.3\times10^{11}$ GeV at 2 $\sigma$ level.
By taking account of uncertainties in the QCD phase transition, 
a conservative constraint would be 
\begin{equation}
  f_a\le 3\times 10^{11}~\mathrm{GeV}.
\end{equation}
Our constraint is a few times tighter than
one given in YKY99. This is partially because of the precise 
$\Omega_\mathrm{CDM}h^2$ obtained from recent observations, which 
was not available at the time of  YKY99, and also of
the axion mass at finite temperature from~\cite{Wantz:2009mi,Wantz:2009it},
which is slightly smaller (hence larger $t_w$) than that adopted in YKY99.

We also note that while the error in $\xi$ of Eq. \eqref{eq:result_xi} 
can be smaller if it is estimated at earlier times $t<t_\mathrm{end}$, 
this does not affect our final consequences much; 
the error in the energy density of axion CDM 
in Eq. \eqref{eq:result_omega} can decrease at most 40\%
and the upper bound on $f_a$ does not change. 

\section{Summary and discussions}\label{sec:summary}

Using field theoretic simulation, 
we have investigated the dynamics of strings and 
the energy spectrum of axion radiation from strings, 
which was previously done by YKY99. Our lattice simulation
has a higher resolution than that of any other previous studies.
We have also introduced several new techniques in order to improve
accuracies of the previous analysis. 
We have developed a new identification 
method of strings in simulation box, which uses the minimum 
phase difference among neighboring grid points in quadrates. The estimated 
scaling parameter $\xi=0.87\pm0.14$ shows good agreement with
other previous studies. The consistency among 
completely different identification methods would suggest
that these results are conclusively robust.

In estimation of the energy spectrum of axions radiated from strings, we
have also introduced a new method to obtain larger statistics.  We have
verified our new method using PPSE, by checking consistency with the
spectrum calculated from sub-boxes in the simulation box found without
strings. The differential energy spectrum obtained from our lattice
simulation peaks sharply around a horizon scales and damps exponentially
toward higher wave numbers. This shows good agreement with YKY99, and
support the discussion of Davis and Shellard \cite{Davis:1989nj}. The
ratio of the horizon scale and the mean energy momentum of radiated
axion is $\epsilon^{-1}=0.23\pm0.02$.  Using these results, we have
obtained a constraint on the axion decay constant $f_a<3\times 10^{11}$
GeV.

Above constraint on $f_a$ counts only axions radiated from strings. 
However, axions are also radiated from axionic domain walls.
This gives a constraint $f_a\lesssim 2.5\times 10^{11}$ GeV \cite{
Hagmann:1990mj, Lyth:1991bb,Nagasawa:1994qu,
Nagasawa:1997zn,Chang:1998tb}, which is also severer than
in the original papers due to the more precise
observationally inferred value of $\Omega_\mathrm{CDM}h^2$.
Moreover, the oscillation of zero mode of axion field also
contributes to the energy density of CDM axions in the present Universe,
which gives $f_a\le (2.8\pm 2)\times 10^{11}$ GeV \cite{Wantz:2009it}. 
Combination of these constraints would give $f_a\lesssim 10^{11}$ GeV, 
as long as there occurs no entropy dilutions after the QCD phase transition.

\bigskip
\bigskip

\noindent 
\section*{Acknowledgment}
We would like to thank Ken'ichi Saikawa for checking analytic calculations.
This work was partially supported by JSPS Grant-in-Aid for Scientific Research
Nos. 19340054 (J.Y.), 2111006 (M.K.), 21244033 (T.H.), 21740187 (M.Y.),
and the Grant-in-Aid for Scientific Research on Innovative Areas
No. 21111006 (M.K. \& J.Y.),
and also supported by World Premier International
Research Center Initiative (WPI Initiative), MEXT, Japan (M.K. \& J.Y.).

\appendix

\section{Unbiasedness of pseudo power spectrum estimator}
\label{app:ppse}
Here we prove that the pseudo power spectrum estimator 
$\hat P(k)$ given in Eq.~\eqref{eq:ppse}
is an unbiased estimator of the energy spectrum $P_\mathrm{free}(k)$
defined in Eq. \eqref{eq:power_free}.
Since $\tilde{\dot a}$ is a convolution of $W$ and $\dot a_\mathrm{free}$, 
the masked spectrum in Eq.~\eqref{eq:masked} can be rewritten as
\begin{equation}
\tilde P(k) =\int \frac{d\Omega_k}{4\pi}\frac{k^2}{V}\int\frac{d^3k^\prime}{(2\pi)^3}
\int\frac{d^3k^{\prime\prime}}{(2\pi)^3} \frac{1}{2}W(\vec k-\vec k^\prime)^*
W(\vec k-\vec k^{\prime\prime})\dot a_\mathrm{free}(\vec k^\prime)^*
\dot a_\mathrm{free}(\vec k^{\prime\prime}).
\end{equation}
Using Eq.~\eqref{eq:power_free}, the ensemble average of the masked spectrum 
can be written in terms of $P_\mathrm{free}(k)$, 
\begin{eqnarray}
\langle\tilde P(k)\rangle&=&
\int \frac{d\Omega_k}{4\pi}\frac{k^2}{V}\int\frac{d^3k^\prime}{(2\pi)^3}\frac{1}{k^{\prime2}}
\left|W(\vec k-\vec k^\prime)\right|^2P_\mathrm{free}(k^\prime) \notag \\
&=&Vk^2\int\frac{dk^\prime}{2\pi^2}M(k,k^\prime)P_\mathrm{free}(k^\prime), 
\label{eq:bias}
\end{eqnarray}
where we have used Eq.~\eqref{eq:winmat} in the second equality.
Unless there are no strings and $W(\vec k)=(2\pi)^3\delta^{(3)}(\vec k)$, $\langle\tilde P(k)\rangle\ne P_\mathrm{free}(k)$.
Therefore $\tilde P(k)$ is in general a biased estimator of $P_\mathrm{free}(k)$.

On the other hand, the ensemble average of PPSE can also be calculated. From Eq.~\eqref{eq:ppse} we obtain
\begin{eqnarray}
\langle \hat P(k)\rangle &=&
\frac{k^2}{V}\int \frac{dk^\prime}{2\pi^2}
M^{-1}(k,k^\prime) \langle\tilde P(k^\prime)\rangle \notag\\
&=&\frac{k^2}{V}\int\frac{dk^\prime}{2\pi^2}M^{-1}(k,k^\prime)
Vk^{\prime2}\int\frac{dk^{\prime\prime}}{2\pi^2}M(k^\prime,k^{\prime\prime})P_\mathrm{free}(k^{\prime\prime})\notag\\
&=&P_\mathrm{free}(k),
\end{eqnarray}
where in the second and third equality, we used Eqs,~\eqref{eq:bias} and \eqref{eq:inverse}, respectively.
Therefore $\hat P(k)$ is an unbiased estimator of $P_\mathrm{free}(k)$. 

\section{Energy density of axion CDM from strings}
\label{app:energy}
The energy density of axions radiated from strings are to be calculated.
When a network of axionic string is in a scaling regime, 
mean energy density of strings is given by
\begin{equation}
\bar\rho_\mathrm{string}(t)=\frac{\xi}{t^2}2\pi f_a^2
\ln\left(\frac{t/\sqrt{\xi}}{d_\mathrm{string}}\right), \label{eq:scaling}
\end{equation}
where in the right hand side, a factor $\xi/t^2$ represents 
the mean density of physical length of strings, 
and the rest $2\pi f_a^2\ln(t/\sqrt{\xi}d_\mathrm{string})$ is the line energy density of strings.
Strings release their energy mainly by emitting massless axions. Therefore
the evolution equations for the system of strings and radiated axions are given by
\begin{eqnarray}
\frac{d\bar\rho_\mathrm{string}(t)}{dt}&=&-2H(t)\bar\rho_\mathrm{string}(t)-
\left[\frac{d\bar\rho_\mathrm{string}(t)}{dt}\right]_\mathrm{emission}, \label{eq:dstring}\\
\frac{d\bar\rho_\mathrm{axion}(t)}{dt}&=&-4H(t)\bar\rho_\mathrm{axion}(t)+
\left[\frac{d\bar\rho_\mathrm{string}(t)}{dt}\right]_\mathrm{emission}.\label{eq:daxion}
\end{eqnarray}
Using Eq.~\eqref{eq:scaling}, the energy loss rate of strings via axion emission is 
obtained from Eq.~\eqref{eq:dstring}, 
\begin{equation}
\left[\frac{d\bar\rho_\mathrm{string}(t)}{dt}\right]_\mathrm{emission}=
2\pi f_a^2\frac{\xi}{t^3}\left[
\ln\left(\frac{t/\sqrt{\xi}}{d_\mathrm{string}}\right)-1\right].
\label{eq:emission}
\end{equation}

We define 
\begin{eqnarray}
\bar N(t)&\equiv& R(t)^3\bar n_\mathrm{axion}(t), \label{eq:number}\\
\bar E(t)&\equiv& R(t)^4\bar \rho_\mathrm{axion}(t).\label{eq:energy}
\end{eqnarray}
$\bar N(t)$ is the number of axions in a unit comoving volume. Both $\bar N(t)$ and
$\bar E(t)$ are constant for massless axions if there are no production or absorption of axions.
Then, Eq.~\eqref{eq:daxion} can be rewritten in terms of $\bar E(t)$. We obtain
\begin{equation}
\frac{dE(t)}{dt}=R(t)^4
\left[\frac{d\bar\rho_\mathrm{string}(t)}{dt}\right]_\mathrm{emission}.
\end{equation}

The number of axions radiated from strings can be given by
\begin{equation}
\bar N(t)=
\int^t_{t_*}dt^\prime
\overline{k^{-1}(t^\prime)}\frac{d\bar E(t^\prime)}{dt^\prime},
\end{equation}
where we have used
\begin{equation}
\frac{d\bar N(t)/dt}{d\bar E(t)/dt}=
\frac{\frac{d}{dt}\left[R(t)^3\int \frac{dk}{2\pi^2}\frac{R(t)}{k}P_\mathrm{free}(k,t)\right]}
{\frac{d}{dt}\left[R(t)^4\int
	      \frac{dk}{2\pi^2}P_\mathrm{free}(k,t)\right]}
=\overline{k^{-1}(t)}.
\end{equation}
$t_*$ is the time when scaling solution is realized. 
Assuming $\overline{k^{-1}(t)}$ is proportional to the  horizon scale 
\begin{equation}
R(t)\overline{k^{-1}(t)}=\frac{1}{\epsilon}\frac{t}{2\pi}, 
\end{equation}
we obtain
\begin{eqnarray}
\bar N(t)&=&
\int^t_{t_*}dt'\frac{t'}{2\pi R(t')}
R(t')^4 2\pi f_a^2\xi\frac{1}{t'^3}\left[
\ln\left(\frac{t'/\sqrt{\xi}}{d_\mathrm{string}}\right)-1\right]\notag\\
&=&\frac{2f_a^2\xi}{\epsilon}
\left[
\frac{R(t')^3}{t'}\left(\ln\left[\frac{t'/\sqrt{\xi}}{d_\mathrm{string}}\right]-3\right)\right]^t_{t_*},
\end{eqnarray}
where in the last line, we have used that $t\propto R(t)^2$ is a good
approximation in the radiation domination. Strictly speaking, we need to
take into account the change of the degree of the freedom of
relativistic particles. However, axion emission from strings continues
until the time of wall formation $t_w\gg t_*$ and actually is dominated
by the contribution at the last time $t_w$. Thus, we can safely omit
such change of the degree of the freedom.  After $t_w$, the number of
axions in a comoving volume is conserved, so that
\begin{equation}
\bar N(t>t_w)=\bar N(t_w)\simeq
2f_a^2\frac{\xi}{\epsilon}\frac{R(t_w)^3}{t_w}
\ln\left(\frac{t_w/\sqrt{\xi}}{d_\mathrm{string}}\right).
\label{eq:naxion}
\end{equation}
The number density of axions from strings are 
$\bar n_\mathrm{axion}(t_0)=\bar N(t_w)/R(t_0)^3$.

Regarding the axion mass at finite temperature,
we quote the recent result from \cite{Wantz:2009mi,Wantz:2009it}, 
where the Interacting Instanton Liquid Model 
is adopted. If a simple power-law fit is applied, the axion mass is given by
\begin{equation}
m_a(T)^2=\alpha\frac{\Lambda^4}{f_a^2}\left(\frac{T}{\Lambda}\right)^{-n},
\label{eq:mass}
\end{equation}
where $n=6.68$, $\alpha=1.68\times10^{-7}$ and $\Lambda=400$~MeV. 
At small $T$, Eq. \eqref{eq:mass} can be arbitrary large, and when it becomes 
larger than the axion mass at zero temperature 
\begin{equation}
m_a(T=0)^2=1.46\times10^{-3}\frac{\Lambda^4}{f_a^2},
\label{eq;mass0}
\end{equation}
we simply set $m_a(T)=m_a(T=0)$. 
This approximation gives good agreement 
with other studies \cite{Bae:2008ue,Turner:1985si}.

At the time $t_w$, the tension of axionic domain wall $\sigma_\mathrm{wall}(t)$ 
begins to dominate the axionic string-wall system, {\it i.e.} 
$\sigma_\mathrm{wall}(t_w)=\mu_\mathrm{string}(t_w)/t_w$.
The wall tension is given by $\sigma_\mathrm{wall}(t)=3\pi m_a(t)f_a^2$.
We assume a constant number of relativistic d.o.f. 
$g_*$ around $t_w$, {\it i.e.} $t_w=1/2H(t_w)$.
Then from Eq. \eqref{eq:mass}, $t_w$ and the temperature $T_w$ at $t_w$ 
are given by
\begin{eqnarray}
t_w&=&2.2\times 10^{-6}~\mathrm{sec} 
\left(\frac{g_{*w}}{70}\right)^{-n/2(n+4)}
\left(\frac{\Lambda}{400\mathrm{MeV}}\right)^{-2}
\left(\frac{f_a}{10^{12}\mathrm{GeV}}\right)^{4/(n+4)},
\label{eq:time_w}\\
T_w&=&0.64~\mathrm{GeV}~
\left(\frac{g_{*w}}{70}\right)^{-1/(n+4)}
\left(\frac{\Lambda}{400\mathrm{MeV}}\right)
\left(\frac{f_a}{10^{12}\mathrm{GeV}}\right)^{-2/(n+4)},
\label{eq:temp_w}
\end{eqnarray}
where $g_{*w}$ is $g_*$ at $t_w$. When hadronic states up to a mass of $3$ GeV
are counted \cite{Wantz:2009it}, we find
$g_{*w}\simeq 70$ for $T_w\simeq 1$ GeV.

Entropy conservation yields
\begin{equation}
R(T_w)^3=7.2\times10^{-40}~\gamma\left(\frac{g_{*w}}{70}\right)
\left(\frac{T_w}{1\mathrm{GeV}}\right)^{-3}, \label{eq:sf_w}
\end{equation}
where $\gamma$ is the dilution factor due to the entropy production after $t_w$.
Finally, combining Eqs. \eqref{eq:naxion}, \eqref{eq:time_w} and \eqref{eq:sf_w},
we obtain the density parameter of axion radiated from axionic strings
\begin{equation}
\Omega_\mathrm{axion}h^2=8.7~\gamma
\frac{\xi}{\epsilon}
\left(\frac{g_{*w}}{70}\right)^{-n/2(n+4)}
\left(\frac{\Lambda}{400\mathrm{MeV}}\right)
\left(\frac{f_a}{10^{12}\mathrm{GeV}}\right)^{(n+6)/(n+4)},
\label{eq:omega_axion}
\end{equation}
where $h$ is the reduced Hubble constant, {\it i.e.} $H_0=100h$~km/sec/Mpc.



\begin{thebibliography}{}

\bibitem{Peccei:1977hh}
  R.~D.~Peccei and H.~R.~Quinn,
  Phys.\ Rev.\ Lett.\  {\bf 38}, 1440 (1977).

\bibitem{Weinberg:1977ma}
  S.~Weinberg,
  Phys.\ Rev.\ Lett.\  {\bf 40}, 223 (1978).

\bibitem{Wilczek:1977pj}
  F.~Wilczek,
  Phys.\ Rev.\ Lett.\  {\bf 40}, 279 (1978).

\bibitem{Raffelt:1990yz}
  G.~G.~Raffelt,
  Phys.\ Rept.\  {\bf 198}, 1 (1990).

\bibitem{Turner:1989vc}
  M.~S.~Turner,
  Phys.\ Rept.\  {\bf 197}, 67 (1990).

\bibitem{inflation}
A.H.\ Guth,
{ Phys.\ Rev.} {\bf D23}, 347 (1981);
K.\ Sato,
{ Mon.\ Not.\ R.\ astr.\ Soc.} {\bf 195}, 467 (1981);
A.A.\ Starobinsky
Phys.\ Lett. {\bf 91B}, 99 (1980).


\bibitem{Wantz:2009it}
  O.~Wantz and E.~P.~S.~Shellard,
  arXiv:0910.1066 [astro-ph.CO].
  
\bibitem{Beltran:2006sq}
  M.~Beltran, J.~Garcia-Bellido and J.~Lesgourgues,
  Phys.\ Rev.\  D {\bf 75}, 103507 (2007).
  [arXiv:hep-ph/0606107].

\bibitem{Kawasaki:2007mb}
  M.~Kawasaki and T.~Sekiguchi,
  Prog.\ Theor.\ Phys.\  {\bf 120}, 995 (2008)
  [arXiv:0705.2853 [astro-ph]].

\bibitem{Kawasaki:2008sn}
  M.~Kawasaki, K.~Nakayama, T.~Sekiguchi, T.~Suyama and F.~Takahashi,
  JCAP {\bf 0811}, 019 (2008)
  [arXiv:0808.0009 [astro-ph]].

\bibitem{Hikage:2008sk}
  C.~Hikage, K.~Koyama, T.~Matsubara, T.~Takahashi and M.~Yamaguchi,
  Mon.\ Not.\ Roy.\ Astron.\ Soc.\  {\bf 398}, 2188 (2009)
  [arXiv:0812.3500 [astro-ph]].

\bibitem{Yokoyama:1988zza}
  J.~Yokoyama,
  Phys.\ Lett.\  B {\bf 212}, 273 (1988).
  J.~Yokoyama,
  Phys.\ Rev.\ Lett.\  {\bf 63}, 712 (1989).
  J.~Yokoyama,
  Phys.\ Lett.\  B {\bf 231}, 49 (1989).

\bibitem{Kawasaki:2010gv}
  M.~Kawasaki, N.~Kitajima and K.~Nakayama,
  arXiv:1008.5013 [hep-ph].

\bibitem{NG}
  D.~P.~Bennett and F.~R.~Bouchet,
  Phys.\ Rev.\  D {\bf 41}, 2408 (1990);
  B.~Allen and E.~P.~S.~Shellard,
  Phys.\ Rev.\ Lett.\  {\bf 64}, 119 (1990).

\bibitem{Moore:2001px}
  J.~N.~Moore, E.~P.~S.~Shellard and C.~J.~A.~Martins,
  Phys.\ Rev.\  D {\bf 65}, 023503 (2002)
  [arXiv:hep-ph/0107171].

\bibitem{Hindmarsh:2008dw}
  M.~Hindmarsh, S.~Stuckey and N.~Bevis,
  Phys.\ Rev.\  D {\bf 79}, 123504 (2009)
  [arXiv:0812.1929 [hep-th]].

\bibitem{Yamaguchi:1998gx}
  M.~Yamaguchi, M.~Kawasaki and J.~Yokoyama,
  Phys.\ Rev.\ Lett.\  {\bf 82}, 4578 (1999)
  [arXiv:hep-ph/9811311].

\bibitem{Yamaguchi:1999yp}
  M.~Yamaguchi,
  Phys.\ Rev.\  D {\bf 60}, 103511 (1999)
  [arXiv:hep-ph/9907506];
  M.~Yamaguchi, J.~Yokoyama and M.~Kawasaki,
  Phys.\ Rev.\  D {\bf 61}, 061301 (2000)
  [arXiv:hep-ph/9910352].

\bibitem{Yamaguchi:2002zv}
  M.~Yamaguchi and J.~Yokoyama,
  Phys.\ Rev.\  D {\bf 66}, 121303 (2002)
  [arXiv:hep-ph/0205308];
  M.~Yamaguchi and J.~Yokoyama,
  Phys.\ Rev.\  D {\bf 67}, 103514 (2003)
  [arXiv:hep-ph/0210343].

\bibitem{Davis:1989nj}
  R.~L.~Davis and E.~P.~S.~Shellard,
  Nucl.\ Phys.\  B {\bf 324}, 167 (1989).

\bibitem{Dabholkar:1989ju}
  A.~Dabholkar and J.~M.~Quashnock,
  Nucl.\ Phys.\  B {\bf 333}, 815 (1990).

\bibitem{Harari:1987ht}
  D.~Harari and P.~Sikivie,
  Phys.\ Lett.\  B {\bf 195}, 361 (1987).

\bibitem{Hagmann:1990mj}
  C.~Hagmann and P.~Sikivie,
  Nucl.\ Phys.\  B {\bf 363}, 247 (1991).
    
\bibitem{Wandelt:2000av}
  B.~D.~Wandelt, E.~Hivon and K.~M.~Gorski,
  Phys.\ Rev.\  D {\bf 64}, 083003 (2001)
  [arXiv:astro-ph/0008111].
    
\bibitem{Hivon:2001jp}
  E.~Hivon, K.~M.~Gorski, C.~B.~Netterfield, B.~P.~Crill, S.~Prunet and F.~Hansen,
  Astrophys.\ J.\  {\bf 567}, 2 (2002)
  [arXiv:astro-ph/0105302].

\bibitem{Hinshaw:2003ex}
  G.~Hinshaw {\it et al.}  [WMAP Collaboration],
  Astrophys.\ J.\ Suppl.\  {\bf 148}, 135 (2003)
  [arXiv:astro-ph/0302217].

\bibitem{Vachaspati:1984dz}
  T.~Vachaspati and A.~Vilenkin,
  Phys.\ Rev.\  D {\bf 30}, 2036 (1984).

\bibitem{Komatsu:2010fb}
  E.~Komatsu {\it et al.},
  [arXiv:1001.4538 [astro-ph.CO]].

\bibitem{Lyth:1991bb}
  D.~H.~Lyth,
  Phys.\ Lett.\  B {\bf 275}, 279 (1992).

\bibitem{Nagasawa:1994qu}
  M.~Nagasawa and M.~Kawasaki,
  Phys.\ Rev.\  D {\bf 50}, 4821 (1994)
  [arXiv:astro-ph/9402066].

\bibitem{Nagasawa:1997zn}
  M.~Nagasawa,
  Prog.\ Theor.\ Phys.\  {\bf 98}, 851 (1997)
  [arXiv:hep-ph/9712341].

\bibitem{Chang:1998tb}
  S.~Chang, C.~Hagmann and P.~Sikivie,
  Phys.\ Rev.\  D {\bf 59}, 023505 (1999)
  [arXiv:hep-ph/9807374].

\bibitem{Wantz:2009mi}
  O.~Wantz and E.~P.~S.~Shellard,
  Nucl.\ Phys.\  B {\bf 829}, 110 (2010)
  [arXiv:0908.0324 [hep-ph]].

\bibitem{Bae:2008ue}
  K.~J.~Bae, J.~H.~Huh and J.~E.~Kim,
  JCAP {\bf 0809}, 005 (2008)
  [arXiv:0806.0497 [hep-ph]].

\bibitem{Turner:1985si}
  M.~S.~Turner,
  Phys.\ Rev.\  D {\bf 33}, 889 (1986).

\end{thebibliography}
\end{document}